\documentclass[doublecol]{epl2} 
\def\I{\openone}
\def\openone{\mathbb I}
\usepackage{bbold}
\usepackage{hyperref}
\usepackage{bbm}
\usepackage{array}
\usepackage{amsmath}
\usepackage{multirow}
\usepackage{amsfonts}
\usepackage{amssymb}
\usepackage{verbatim}
\usepackage{leading}

\title{Spin-half bosonic classification}
\shorttitle{Spin-half bosonic classification} 

\author{R. J. Bueno Rogerio\inst{1} \and A. R. Aguirre\inst{1}}
\shortauthor{R. J. Bueno Rogerio \etal}

\institute{                    
  \inst{1} Institute of Physics and Chemistry, Federal University of Itajub\'a , Itajub\'a, Minas Gerais, 37500-903, Brazil.
}

\abstract{
In this paper, we define a new spinor classification that encompasses the recently proposed spin-half bosons with mass dimension three-half \cite{dharamboson}. As it will be shown, these particles, which are governed by a first-order equation and consequently provide a local theory, belong to a specific subclass of class-2 spinors within the bosonic classification, playing a similar role to Dirac spinors within the Lounesto classification.
Such bosonic classification is shown to be closely connected with the usual fermionic Lounesto classification, and thus, evincing a symmetry between Dirac fermions and spin-half bosons.}

\begin{document}

\maketitle

\section{Introduction}\label{intro}

A series of interesting events, like the recent theoretical discovery of new spin one-half fermions endowed with mass dimension one, provides a strong reason to believe that the fundamental structure of the Quantum Field Theory (QFT) are certainly being improved and strengthened \cite{mdobook,dharamnewfermions,tipo4epjc,chengtipo42021}. 

Recently, it has been proposed a new different type of unification of fermions (more precisely Dirac fermions) and bosons \cite{dharamboson}.  Unlike the supersymmetric field theoretical formalism, where the unification of bosonic and fermionic fields is encompassed within a superfield residing in a Grassmaniann extension of the ordinary Minkowski space, a.k.a. superspace \cite{Wess,Srivastava}, in this recent proposal the new bosonic spinors  and the standard Dirac spinors belong to the same irreducible representation of the Lorentz group.

Following those ideas, in this work we then deal with a entirely new class of spin-half bosons that reside in the $(1/2,0)\oplus(0,1/2)$ representation space, endowed with mass dimensionality three-half rather than one, and describe a complete set of eigenspinors of parity operator, just like the Dirac spinors. Such new particles belong to a subclass of class-2 spinors within a specific bosonic classification, which can provide us a hint towards dynamics and quantum field locality \cite{rjfermionicfield}. These new spinors will then emerge from some (specific) set of arrangement of the Clifford algebra basis, by using the well-known linearly independent sets of the square roots of the $4\times 4$ identity matrix ($\I$) --- the set of matrices $\Gamma_j$ \cite{schweber}.

On the other hand, the well-known Lounesto's spinor classification usually classify spin-half fermions accordingly their bilinear invariants (physical information) \cite{lounestolivro}.
In that classification, the Dirac spinors are defined as the spinor whose scalar and pseudo-scalar bilinear quantities do not vanish simultaneously. Nonetheless, this classification also encompasses different type of spinors carrying both scalar and pseudo-scalar bilinear quantities  which does vanish identically. Those spinors are usually called singular spinors.

It turns out that the spin-half bosons are defined in a bosonic sector of Clifford algebra, and then it is interesting to study the bosonic counterpart of Lounesto's classification, by investigating at first the bilinear forms and their properties for a better understanding of these new particles. 
Up to our knowledge, usually mass dimension one fermions are encompassed within what is known as \emph{Beyond the Standard Model} (BSM) of particle physics --- such as Elko spinors \cite{mdobook}, flag-dipole spinors \cite{tipo4epjc} and the dipole spinors \cite{dharamnewfermions,rodolfodipole}. Accordingly, a deep study of the main properties of the recent particles proposed in \cite{dharamboson}, certainly will improve our comprehension about what the particle content of the BSM could be.

The manuscript is organized as follows: the next section is reserved for a brief overview on the bosonic representation of the Clifford algebra. Posteriorly, we analyse the main aspects of the spin-half bosons bilinear densities looking towards define a bosonic classification. Going further, we connect our results with the ones already established for the fermionic Dirac case --- evincing a strong symmetry among them. Finally, we present some concluding remarks.

\section{On the bosonic representation of the Clifford algebra}\label{cliffordboson}
The usual constitutive relation of the Clifford algebra is given by
\begin{eqnarray}\label{clifffermion}
\lbrace \gamma_{\mu}, \gamma_{\nu}\rbrace = 2g_{\mu\nu}\I, \quad \mu,\nu = 0,1,2,...,N-1,
\end{eqnarray}
where $g_{\mu\nu}$ is a $N=2n$ even-dimensional space-time metric, which in Cartesian coordinates has the form $diag(1,-1,...,-1)$. The generators of the Clifford algebra are the identity $\I$ and the vectors $\gamma_\mu$, usually represented as square matrices, which in the Weyl representation, read
\begin{eqnarray}\label{gammamatrices}
\gamma_0 = \left( \begin{array}{cc}
0 & \I \\ 
\I &0
\end{array}\right), \qquad \boldsymbol{\gamma} = \left( \begin{array}{cc}
0 & \boldsymbol{\sigma} \\ 
-\boldsymbol{\sigma} & 0
\end{array}  \right),
\end{eqnarray}
where $\boldsymbol{\sigma}$ stand for the Pauli matrices in the standard representation. The introduced set of $\gamma$'s allow one to define
\begin{eqnarray}
\gamma = \frac{i}{4!}\epsilon^{\mu\nu\alpha\beta}\gamma_{\mu}\gamma_{\nu}\gamma_{\alpha}\gamma_{\beta} = \left(\begin{array}{cc}
\I & 0 \\ 
0 & -\I
\end{array} \right),
\end{eqnarray}
where $\epsilon^{\mu\nu\alpha\beta}$ is the completely antisymmetric fourth rank tensor with $\epsilon^{0123}=1$.
Such an algebra is necessary to fully describe the geometric structure of the space of physical observable quantities of spin-$1/2$ particles \cite{crawford1}.

Now, following the construction previously introduced in \cite{dharamboson}, we investigate the possibility to define a new set of basis elements of the Clifford algebra. We look towards investigating the Clifford algebra's constitutive relations, and employ the Dirac normalization \cite{crawford1} to define bispinorial densities.  
 
Thus, we introduce a new set of matrices which, in the Weyl representation  \cite{dharamboson}, read
\begin{eqnarray}
a_0 = i\left(\begin{array}{cc}
0 & \I \\ 
-\I & 0
\end{array}\right), \quad \boldsymbol{a} = i\left(\begin{array}{cc}
0 & \boldsymbol{\sigma} \\ 
\boldsymbol{\sigma} & 0
\end{array} \right), 
\end{eqnarray}
satisfying the following properties: $a^2_0=\I$, $\boldsymbol{a}^2=-\I$, and $\boldsymbol{a}^{\dag}=a_0\boldsymbol{a}a_0$. Moreover, they satisfy a very similar constitutive relation as displayed in \eqref{clifffermion}, standing for the bosonic counterpart for the Clifford algebra \cite{brauer}
\begin{eqnarray}\label{cliffboson}
\lbrace a_{\mu}, a_{\nu}\rbrace = 2g_{\mu\nu}\I.
\end{eqnarray}
and also they  do not commute with $\gamma_{\mu}$,
\begin{eqnarray}
[a_{\mu},\gamma_{\nu}] = 2g_{\mu\nu}i\gamma.
\end{eqnarray}
In addition, we can also define
\begin{eqnarray}\label{a5}
a = -\frac{i}{4!}\epsilon^{\mu\nu\alpha\beta}a_{\mu}a_{\nu}a_{\alpha}a_{\beta} = -\left(\begin{array}{cc}
\I & 0 \\ 
0 & -\I
\end{array} \right).
\end{eqnarray}
From these definitions, we have the following relations,
\begin{eqnarray}\label{a-gamma}
a_{\mu} = i\gamma\gamma_{\mu}, \quad \gamma_{\mu} = i a a_{\mu}.
\end{eqnarray} 
Such relations allow one to connect the fermionic with the bosonic representation of the Clifford algebra. Interestingly enough, a fact that deserves our attention, is that the very definition presented in Eq.\eqref{a-gamma} ensures the possibility of interchanging the fermionic and bosonic statistics of the fields, i.e., a new symmetry encoded in the aforementioned equation allows one to transmute from fermionic into bosonic fields. The main reason why this happens is not completely clear to the authors at the moment. Nonetheless, it is worth pointing out that the chirality symmetry generator, namely $a$, does generate a conserved axial current in the bosonic sector (even in the massive case), since the spin-half bosons belong to class-2 spinors ($\omega_{B} = 0$), as it will be shown in the next sections.

\section{On the spin-half bosonic bilinear forms}\label{bosoniclassification}

The usual approach requires a complementation of the (fermionic) Clifford algebra basis, in order to ascertain real bilinear densities \cite{crawford1}. Thus, we now introduce a bosonic complement given by the composition of the $a_{\mu}$ vector basis, i.e.
\begin{eqnarray}\label{atil}
\tilde{a}_{\mu_{1}\mu_{2}...\mu_{N-M}} \equiv \frac{1}{M!}\epsilon_{\mu_{1}\mu_{2}...\mu_{N}}a^{\mu_{N-M+1}}a^{\mu_{N-M+2...}}a^{\mu_{N}}.
\end{eqnarray} 
Bearing in mind we are working in $(1+3)$ dimensional space-time ($N=4)$, the lowest value for $M$ is 2 (the smallest basis arrangement), and then we have that $M=2,3,4$. Now, by analogy with the prescription used by Crawford in \cite{crawford1}, we define the elements of the bosonic counterpart of Clifford algebra basis, as follows
\begin{eqnarray}\label{set}
\lbrace A_{i}\rbrace \equiv \lbrace \I, a, a_{\mu}, aa_{\mu}, a_{\mu\nu} \rbrace,  
\end{eqnarray} 
with $a$ previously defined in \eqref{a5}.    

Looking towards define all the bi-spinor densities encoded on the set \eqref{set}, firstly, we investigate the primordial two bilinears arising from the Clifford algebra basis ($\I$ and $a_{\mu}$), described by 
\begin{eqnarray}
\sigma &=&\bar{\lambda}\lambda,\label{sig}\\
J_{\mu} &=&\bar{\lambda} a_{\mu}\lambda\label{jota},
\end{eqnarray}
where $\bar{\lambda}=\lambda^{\dag}\eta$ is identified as the standard dual spinor, with $\eta$  a $4\times 4$ matrix. From demanding reality conditions, 
\begin{eqnarray}
\sigma = \sigma^{\dag},
\\
J_{\mu} = J_{\mu}^{\dag},
\end{eqnarray}
we find that $\eta^{\dag}=\eta$, and $\eta^{-1}a_{\mu}^{\dag}\eta=a_{\mu}$. Therefore, these lead us to the following solution, $\eta=a_0$. Thus, the bosonic adjoint structure reads $\bar{\lambda}=\lambda^{\dag}a_0$.  It is also worth noting the following relations\footnote{Notice the same commutative/anti-commutative relations are accomplished by requiring the replacement $a_0=i\gamma\gamma_0$ or $\gamma_0=ia a_{0}$, showing a strong connection among $a_{\mu}$ and $\gamma_{\mu}$ matrices. }  
\begin{eqnarray}\label{lorentzgenerators}
\{a_0,\boldsymbol{\kappa}\}=0, \quad [a_0,\boldsymbol{\zeta}]=0,
\end{eqnarray}
where $\kappa$ and $\zeta$ stand for the boost and rotation generators in the $(0,1/2)\oplus(1/2,0)$ representation, which in its matrix form read
\begin{eqnarray}
\boldsymbol{\kappa} = \left(\begin{array}{cc}
-i\boldsymbol{\sigma}/2 & 0 \\ 
0 & i\boldsymbol{\sigma}/2
\end{array} \right), \quad \boldsymbol{\zeta}=\left(\begin{array}{cc}
\boldsymbol{\sigma}/2 & 0 \\ 
0 & \boldsymbol{\sigma}/2
\end{array} \right). 
\end{eqnarray}
Such relations ensure a consistent dual spinor definition and a Lorentz invariant norm. Note that for the usual Dirac fermionic set-up, one must replace $a_0$ by $\gamma_0$. 

We may now proceed computing the remaining arrangements of the basis elements in order to define the bilinear covariants. Taking into account Eq. \eqref{atil} and the requirement of real bispinor densities, 
the basis elements are given, accordingly, by 
\begin{eqnarray}
\tilde{a} = i a \qquad (M=4), \\
\tilde{a}_{\mu} = -a a_{\mu} \qquad(M=3),\\
\tilde{a}_{\mu\nu} = -\frac{i}{2}a_{\mu}a_{\nu}\qquad (M=2).
\end{eqnarray}
Now, with the bosonic representation of the Clifford algebra at hands, it is possible to construct the following bilinear forms, given by
\begin{eqnarray}\label{bosonbilinear}
&&\sigma_{B} = \lambda^{\dag}a_0\lambda,  \\
&& \omega_{B} = i\lambda^{\dag}a_0 a\lambda,\\
&& \textbf{J}_{B}= \lambda^{\dag}a_0 a_{\mu}\lambda\; \theta^{\mu},  \\
&& \textbf{K}_{B}= -\lambda^{\dag}a_0aa_{\mu}\lambda\; \theta^{\mu}, \\
&& \textbf{S}_{B} = -i\lambda^{\dag}a_0 a_{\mu}a_{\nu}\lambda\; \theta^{\mu}\wedge \theta^{\nu},\label{bosonbilinear25}
\end{eqnarray}
where the label $B$ stands for bosonic, and the elements $\{ \theta^\mu \}$ are the dual basis of a given inertial-frame $\{ \textbf{e}_\mu \} = \left\{ \frac{\partial}{\partial x^\mu} \right\}$, with $\{x^\mu\}$ being the global space-time coordinates. 
The above prescription is quite important, since it will guarantee the right appreciation of the physical information encoded in the spin-half bosons.

The bilinear forms introduced in eqs. (\ref{bosonbilinear})--(\ref{bosonbilinear25}) ensure that the Fierz-Pauli-Kofink identities \cite{lounestolivro}, are satisified. Therefore, we now are able to define a bosonic counterpart of the Lounesto's classification
\begin{enumerate}
  \item[$1_{B})$] $\sigma_{B}\neq0$, $\quad \omega_{B}\neq0$, \quad  $\textbf{K}_{B}\neq 0,$ \quad $\textbf{S}_{B}\neq0$;
  \item[$2_{B})$] $\sigma_{B}\neq0$, $\quad \omega_{B}=0$, \quad $\textbf{K}_{B}\neq0,$ $\quad\textbf{S}_{B}\neq0$;
  \item[$3_{B})$] $\sigma_{B}=0$, $\quad \omega_{B}\neq0$, \quad $\textbf{K}_{B}\neq0,$ $\quad\textbf{S}_{B}\neq0$;
  \item[$4_{B})$] $\sigma_{B}=0=\omega_{B},$ \hspace{0.98cm}  $\textbf{K}_{B}\neq0,$ $\quad\textbf{S}_{B}\neq0$;
  \item[$5_{B})$] $\sigma_{B}=0=\omega_{B},$ \hspace{0.98cm} $\textbf{K}_{B}=0,$ $\quad\textbf{S}_{B}\neq0$;
  \item[$6_{B})$] $\sigma_{B}=0=\omega_{B},$ \hspace{0.98cm} $\textbf{K}_{B}\neq0,$ $\quad\textbf{S}_{B}=0$;
\end{enumerate}  
where we notice that the bilinear $\textbf{J}_{B}$ is always non-null for all classes. 

Since the  matrices $a_{\mu}$ are incorporated in a bosonic field theoretic structure \cite{brauer}, we regard our representation as a bosonic representation of the Lounesto's classification, in which we find the possibility of defining six classes of bosons.
A careful inspection of Eq. \eqref{a-gamma} evince a new symmetry between fermions and spin one-half bosons --- which brings the possibility to interchange the bosonic bi-spinorial densities into the fermionic bi-spinorial densities, and vice-versa. This symmetry may provide the equivalent bosonic of the spinors defined within  Lounesto classification. Then, we should suppose the existence, within the regular sector, of a spin-half boson that holds a similar properties to the Dirac spinors. Whereas, in the singular sector, it is very likely to find bosons holding similar features to Majorana (flag-pole), flag-dipole, and dipole spinors.

\section{Remarks on \emph{class-2} spin-half bosons}
In this section we call attention for some details concerning the particles introduced in \cite{dharamboson}. The bosons residing in the $(1/2, 0) \oplus (0, 1/2)$ representation space, under a judicious choice of the phases factors, reads
\begin{eqnarray}
\lambda_1(\textbf{0}) = \sqrt{\frac{m}{2}}\left(\begin{array}{c}
-1 \\ 
i \\ 
i \\ 
1
\end{array}\right),\; \lambda_2(\textbf{0}) = \sqrt{\frac{m}{2}}\left(\begin{array}{c}
1 \\ 
i \\ 
-i \\ 
1
\end{array}\right),
\end{eqnarray} 
and
\begin{eqnarray}
\lambda_3(\textbf{0}) = \sqrt{\frac{m}{2}}\left(\begin{array}{c}
1 \\ 
-i \\ 
i \\ 
1
\end{array}\right),\; \lambda_4(\textbf{0}) = \sqrt{\frac{m}{2}}\left(\begin{array}{c}
-1 \\ 
-i \\ 
-i \\ 
1
\end{array}\right).
\end{eqnarray} 
To define the spinors for an arbitrary momentum, $\lambda_{j}(\textbf{p})=\kappa\lambda_{j}(\textbf{0})$, it is sufficient to act with the $(1/2, 0)\oplus(0, 1/2)$ boost operator
\begin{eqnarray}
\kappa = \sqrt{\frac{E+m}{2m}}\left(\begin{array}{cc}
\I+ \frac{\vec{\sigma}\cdot\vec{\textbf{p}}}{E+m} & 0 \\ 
0 & \I- \frac{\vec{\sigma}\cdot\vec{\textbf{p}}}{E+m}
\end{array} \right).
\end{eqnarray} 
Under the adjoint structure $\bar{\lambda}(\textbf{p}) = \lambda^{\dag}(\textbf{p})a_0$, it provides the Lorentz-invariant norms
\begin{eqnarray}
&&\bar{\lambda}_j(\textbf{p})\lambda_j(\textbf{p}) = +2m, \quad\mbox{for j = 1, 2},
\\
&&\bar{\lambda}_j(\textbf{p})\lambda_j(\textbf{p}) = -2m, \quad\mbox{for j = 3, 4}.
\end{eqnarray} 
The above spinors satisfy the following dynamical equation
\begin{eqnarray}
(a_{\mu}p^{\mu}-m\I)\lambda_j(\textbf{p}) = 0, \quad\mbox{for j = 1, 2}, \label{diracboson1}
\\
(a_{\mu}p^{\mu}+m\I)\lambda_j(\textbf{p}) = 0, \quad\mbox{for j = 3, 4}.\label{diracboson2}
\end{eqnarray}
Note that the equations \eqref{diracboson1} and \eqref{diracboson2} stands for the Dirac counterpart for the bosonic case.
The previous observation allows us to state that such bosons are classified as class-2 within the introduced bosonic classification, and 
thus, we can infer that such bosons necessarily provide a local theory, similarly as the discussions in \cite{rjfermionicfield} around the Dirac class-2 spinors. With these observations in mind, we now may entrain on a concrete example culminating in the new local bosonic fields of spin one-half.
 
Recalling the definition of the parity operator $\mathcal{P}=m^{-1}\gamma_{\mu}p^{\mu}$ \cite{speranca}, now, we may define the bosonic parity operator. With relations \eqref{a-gamma} and the equations \eqref{diracboson1} and \eqref{diracboson2} at hands, we may define
\begin{equation}
\mathcal{P}_{B} = m^{-1}a_{\mu}p^{\mu}
\end{equation}
or
\begin{equation}
\mathcal{P}_{B} = i\gamma\mathcal{P}.
\end{equation}
The operators $\mathcal{P}_{B}$ and $\mathcal{P}$ are connected, showing explictly the connection among the fermionic and the bosonic sectors, where the $\gamma$ matrix play a central role.
A simple exercise reveal us
\begin{eqnarray}
&&\mathcal{P}_{B}\lambda_1(\textbf{p})=\lambda_1(\textbf{p}), \quad\,\,\,\, \mathcal{P}_{B}\lambda_2(\textbf{p})=\lambda_2(\textbf{p}),
\\
&&\mathcal{P}_{B}\lambda_3(\textbf{p})=-\lambda_3(\textbf{p}), \quad \mathcal{P}_{B}\lambda_4(\textbf{p})=-\lambda_4(\textbf{p}).
\end{eqnarray} 
Standing for a class of eigenpinors of the parity operator, with the consequence that $\mathcal{P}^2_{B} = +\I$ --- as Dirac spinors do. For completeness, we highlight the behaviour of $\lambda_j(\textbf{p})$ under the other discrete symmetries $\mathcal{C}^2=+\I$ and $\mathcal{T}^2=-\I$, consequently, we obtain $(\mathcal{C}\mathcal{P}\mathcal{T})^2=+\I$ \cite{dharamboson}, in which, $\mathcal{C} = \gamma_2\mathcal{K}$ (where $\mathcal{K}$ stands for the algebraic complex conjugation operation) and $\mathcal{T} = i\gamma\mathcal{C}$, being the charge-conjugation and time-reversal operator, respectively.

\section{Concluding Remarks}\label{remarks}
The focus of the present paper is to show some details of a quite recent class of spin-half bosons and advance in a suitable classification for such particles. The right appreciation of the bosonic sector of the Clifford algebra is a mechanism that makes possible to accomplish given task. An interesting feature encoded in the present  results lies on the \emph{symmetry} between the Dirac fermions (which can also be extended to other fermions) and spin-half bosons, see Eq. \eqref{a-gamma}. Such feature allows to transmute among fermionic and bosonic sectors of the Clifford algebra.

At this point, it should be interesting made some comments about the differences between this symmetry and supersymmetry. It is well-known that within the context of supersymmetric field theories the unification of bosonic and fermionic fields is accomplished by introducing a superfield residing in a Grassmaniann extension of the ordinary Minkowski space, which is named superspace. The symmetry between the fields is manifestly implemented through the supersymmetry generators, which satisfy anticommutation relations and extent the spacetime symmetry algebra to the so-called super-Poincaré algebra. On the contrary, in the approach we are following, no extension of the spacetime algebra is needed, since Dirac fermions and the new bosons live in the same irreducible representation of the Lorentz symmetry algebra, and  we should rather regard this new symmetry as an unification of both sort of spinors by identifying the corresponding representations of the Clifford algebra. We believe that the study of the corresponding supersymmetric construction based on these new bosons deserves a deeper analysis, which we expect to address it soon in future works.

So far we have evidence of only one class of spin-half boson, the one introduced in this work belong to class-2, and in view of this fact, such particle hold similar properties to Dirac fermions, i.e. satisfy a first-order equation, form a complete set of eigenspinors of parity operator and, in addition, provide a local theory. We suppose that from the mechanism introduced in \cite{dharamboson}, it may be possible to find more spin-half bosons, which in that case might hold properties similar to the eigenspinors of the charge-conjugation operator, Majorana, and dipole fermions.

Another interesting aspects is that, according to the Shirokov-Trautman classification \cite{shirokov,trautman}, spinors may be classified scrutinizing all the possible signs $\lambda, \mu, \nu \in \{+,- \}$ in the following relations: $\mathcal{P}\mathcal{T} = \lambda\mathcal{T}\mathcal{P}$,  $\mathcal{P}^2 = \mu\I$,  $\mathcal{T}^2 = \nu\I$. Commutativity of $\mathcal{P}$ and $\mathcal{T}$ gives $\lambda= +$, while the other results provides $\mu = +$ and $\nu = - $. Thus, spin-half boson belongs to a class identified as $(\lambda,\mu,\nu) = (+,+,-)$, similarly as mass dimension one fermions in \cite{mdobook}.
Moreover, according to the Wigner classification, spin-half bosons belong to the same class as Elko spinors. More detailed investigations must be provided.

It would be also interesting to seek for further development, in the light of \cite{rodolfoconstraints}, looking for more general solutions for spin-half bosons, as well as finding out a suitable way to map fermionic into bosonic propagators via bilinear densities. These issues are already under investigation.

\bibliographystyle{unsrt}
\bibliography{refs}

\end{document}